\documentclass[
reprint,
nofootinbib,
aps, prl
]{revtex4-1}
\usepackage{custom}
\usepackage{comment}
\usepackage{float}
\usepackage{graphicx} 
\usepackage[caption=false]{subfig}
\usepackage{float}
\usepackage{amsmath, amsfonts}
\usepackage{MnSymbol}
\usepackage{bm}	
\usepackage{soul}
\usepackage{comment}
\usepackage{bbm}				 
\usepackage{braket}				 
\usepackage{mathtools}			 
\usepackage{graphicx} 			 
\usepackage{pst-plot}			 
\usepackage[hidelinks]{hyperref} 
\usepackage{color}
\usepackage{CJKutf8}

\renewcommand{\Re}{\mathop{\rm Re}}		
\renewcommand{\Im}{\mathop{\rm Im}}		

\begin{document}

\title{Optomechanical quantum entanglement mediated by  acoustic phonon fields}

\author{Qidong Xu}
\email[]{qidong.xu.gr@dartmouth.edu}

\author{M. P. Blencowe}
\email[]{miles.p.blencowe@dartmouth.edu}

\address{Department of Physics and Astronomy, Dartmouth College, Hanover, New Hampshire 03755, USA}

\date{\today}

\begin{abstract}
We present exact solutions for the quantum time evolution of two spatially separated, local inductor-capacitor (LC) oscillators that are coupled optomechanically to a long elastic strip that functions as a quantum thermal acoustic field environment. We show that the optomechanical coupling to the acoustic environment gives rise to causal entanglement dynamics between the two LC oscillators in the absence of resonant photon exchange between them, and that significant entanglement develops 
regardless of the environment temperature. Such a process establishes that distributed entanglement may be generated between superconducting qubits via a connected phonon bus bar,  without the need for resonant phonon release and capture.
\end{abstract}

\maketitle

{\it Introduction.---} Thermal environments have often been invoked to explain the decoherence of a quantum system, thus  resulting in the observed classical, macroscopic world \cite{zurek1982environment,joos1985emergence,zurek1991quantum}. However, it is also well known that thermal environments can generate quantum entanglement when coupled to otherwise independent quantum subsystems under suitable conditions \cite{braun2002creation,benatti2003environment,oh2006entanglement,romano2006environment,ferreira2006macroscopic,contreras2008entanglement,paz2008dynamics,mccutcheon2009long,zell2009distance,galve2010bringing,eastham2016bath,hu2018steady}; several experimental realizations have been proposed \cite{braun2002creation,retzker2005detecting,sabin2010dynamics,sabin2012extracting,cattaneo2021bath}, with further examples considered in the  Ref. \cite{aolita2015open} review (and references therein).

In this Letter, we investigate the entanglement dynamics of an experimentally feasible model comprising two spatially separated inductor-capacitor (LC) oscillators that are capacitively coupled to a long, partially metallized elastic strip via the optomechanical interaction \cite{aspelmeyer2014cavity}; here, the elastic strip functions as a thermal acoustic phonon environment. Hybrid quantum information platforms with acoustic phonons serving as the mediators have received increasing attention in recent years \cite{bienfait2019phonon,mirhosseini2020superconducting,dumur2021quantum,zivari2022chip}, in part due to their long coherence times \cite{maccabe2020nano} and much lower propagation speeds compared with photons \cite{delsing20192019}. While most studies have focused on single mode resonant phonon dynamics models, we instead adopt a field theoretic description of the elastic strip in our model, which naturally leads to local,  spatially-dependent non-resonant couplings between the oscillators and the phonon field. This then allows for an explicit analysis of the causal nature of the entanglement dynamics between the two oscillators arising  from the finite acoustic wave propagation  speed in the elastic strip. Tracing out the elastic strip (phonon) degrees of freedom, we solve {\it exactly} for the quantum time evolution of the LC oscillators, with particular attention paid to the competing entanglement and dephasing/rephasing dynamics of the LC oscillators. In particular, we find that the two LC oscillators can become substantially entangled  due to their couplings to the much lower frequency acoustic phonon modes, which can be engineered to have significantly low transmission loss rates \cite{ghadimi2018elastic}.

With the capacitor sizes much smaller than the elastic strip length, the two LC oscillators can also be thought of as variants of the so-called  Unruh-DeWitt (UDW) photon detector model \cite{unruh1976notes,dewittGeneralRelativityEinsten1979,lin2007backreaction}; we find that the entanglement only forms between the two LC oscillators when they are `timelike' separated (i.e., causally connected), as opposed to `spacelike' separated, with respect to the acoustic wave propagation (i.e., phonon)  speed. This is to be contrasted with the results obtained for the usually considered bilinear type interaction between the UDW detectors and the field, where entanglement can be `harvested' from the field vacuum state even for spacelike separated detectors \cite{reznik2003entanglement,reznik2005violating,lin2010entanglement,martin2013sustainable,salton2015acceleration}. Such a difference lies in the fact that the optomechanical interaction commutes with the free Hamiltonian of the LC oscillators, and therefore obeys the general no-go theorem of Ref. \cite{simidzija_general_2018} for entanglement generation when the two detectors are `spacelike' separated. 

The optomechanical interaction bears some similarities with the weak field, scalar matter-graviton interaction action \cite{blencowe2013effective,xu2020toy}. In particular, our model Hamiltonian [see Eq.~(\ref{Hamiltonian})] takes on the same form as considered in recent proposals \cite{bose2017spin,marletto2017gravitationally} to observe quantum gravity induced entanglement at low energies \cite{remarks}. The
model can therefore serve as a gravitational entanglement
generation analog to explicitly inform how the mediating field is responsible for the entanglement generation, in contrast to these proposals where only the effective Newtonian potential was considered.
\begin{figure}[h]
\begin{center}
\includegraphics[width=3.5in]{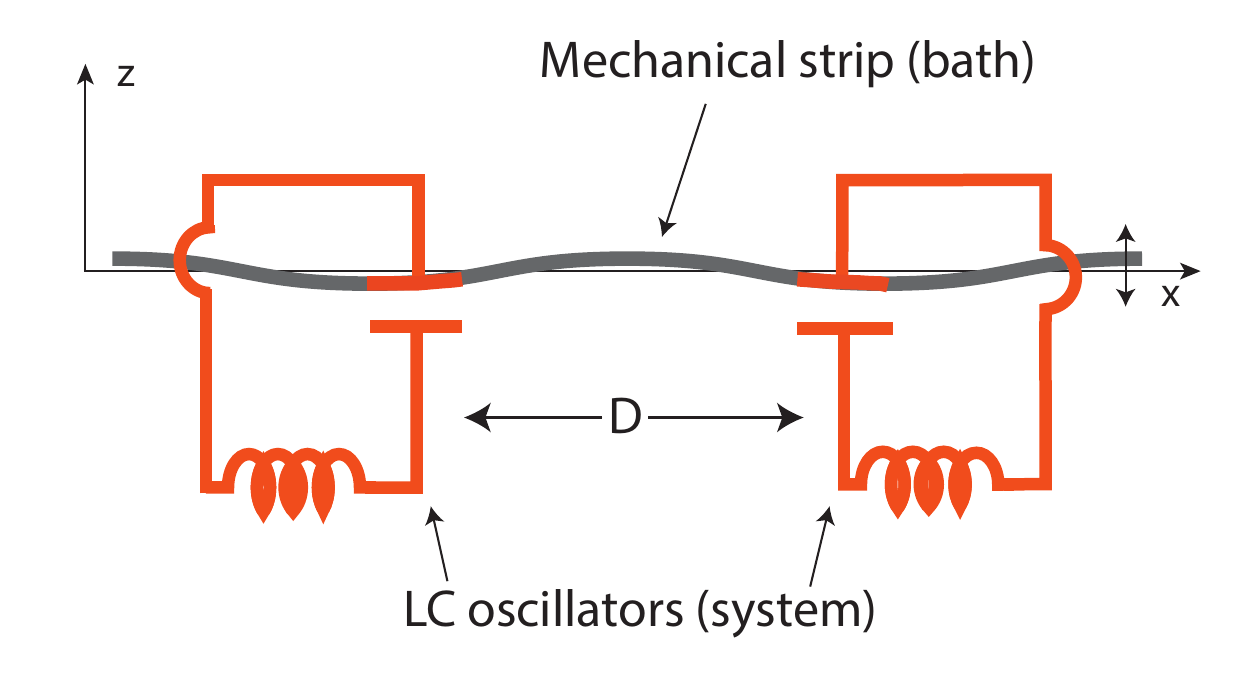} 
\caption{\label{fig1dmodel}  Scheme of the model system. Two spatially separated LC circuit oscillators (system) are capacitively  coupled  to a long oscillating, elastic strip (environment) via two metallized segments.}
\end{center}
\end{figure}

{\it The model.---} Our model scheme (Fig.~\ref{fig1dmodel}) builds on the one considered in Ref. \cite{xu2021cavity}, which investigated dephasing only of a single LC oscillator  coupled capacitively to a long elastic strip. In particular, we consider two identical LC circuits separated by a distance $D$, each coupled capacitively via metallized segments (with lengths $\Delta L$) of a long, elastic mechanical strip with overall length $L>D\gg\Delta L$ that is clamped at both ends. The LC circuits are sited  such that the center point between the two capacitors coincides with the strip center. The transverse width ($W$) and thickness ($T$) of the strip satisfy $T\ll W\lll L$. The indicated lower capacitor plates are assumed fixed, also with length $\Delta L$, the same width $W$ as the strip, and separated from the upper flexing, metallized $\Delta L$ strip  segments of the  strip by a small equilibrium vacuum gap  $d \ll W$. The bare, zero flexing capacitance of each LC circuit is then given by the standard parallel plate expression ${\mathsf{C}}_b = \epsilon_0 W \Delta L/d$ with $\epsilon_0$ the vacuum permittivity. In the following we shall denote the left circuit capacitance by $\mathsf{C}_l$ and right circuit capacitance by $\mathsf{C}_r$, and we denote both circuit inductances by ${\mathsf{L}}$. 

Neglecting displacements in the transverse $y$ and longitudinal $x$ directions, the flexing mechanical displacement of the strip along the transverse $z$ direction can be described by the Hamiltonian
\begin{align}
\mathcal{H}_{\mathrm{bath}} = \frac{\rho_m W T}{2} \int_{0}^{L} dx \left(\frac{\partial u_z}{\partial t}\right)^2 + \frac{F}{2} \int_0^{L} dx \left(\frac{\partial u_z}{\partial x} \right)^2,
\label{hameq}
\end{align}
where $u_z(t,x)$ is the displacement field, $\rho_m$ is the mass density of the strip, and we assume a sufficiently large tensile force $F$ is applied at both ends of the strip so that it behaves effectively as a string with end boundary conditions $u_z(t,x = 0) = u_z(t,x = L) = 0$.

The Hamiltonian for the  two LC circuit system is
\begin{align}
 \mathcal{H}_{\mathrm{sys}} = \frac{Q_l^2}{2\mathsf{C}_l} + \frac{\Phi_l^2}{2\mathsf{L}} + \frac{Q_r^2}{2\mathsf{C}_r} + \frac{\Phi_r^2}{2\mathsf{L}},
\end{align}
where $Q$ is the capacitor charge coordinate and $\Phi$ is the inductor flux coordinate with subscript $l$ and $r$ denoting left and right circuit, respectively. We note that $\mathsf{C}_l$ and $\mathsf{C}_r$ are implicit functions of the displacement field $u_z(t,x)$, with $\mathsf{C}_l(u_z = 0) = \mathsf{C}_r(u_z = 0) \equiv C_b$.

Introducing  creation/annihilation operators for both the LC circuits and the elastic strip modes, and expanding the LC circuit resonant frequencies and creation/annihilation operators to first order in the strip transverse displacement field with the usual rotating wave approximations, the total Hamiltonian reduces to the standard optomechanical Hamiltonian \cite{xu2021cavity}
\begin{align}
\mathcal{H}  =& \sum_{k=1}^2 \Big[ \hbar \Omega_b \left(a_k^\dag a_k +\frac{1}{2}\right) +\sum^\infty_{j = 1} \hbar g_{k,j} \left(a_k^\dag a_k +\frac{1}{2}\right)   \left(b_j +b_j^\dag\right) \Big] \nn \\
&+ \sum^\infty_{j=1} \hbar \omega_j \left(b_j^\dag b_j +\frac{1}{2} \right),
\label{Hamiltonian}
\end{align}
where $a_k$ ($a_k^\dag$) are the annihilation (creation) operators for the LC oscillators with bare frequency $\Omega_b = 1/\sqrt{\mathsf{C}_b \mathsf{L}}$, with the subscript $k=1\, (2)$ denoting respectively the left (right) LC oscillator, and $b_j$ ($b_j^\dag$) are the annihilation (creation) operators for the elastic strip modes of frequency $\omega_j = \pi j \sqrt{\frac{F}{2 m L}}$, with $m=\rho_m W T L/2$ the effective mass of the modes. The coupling strength between each LC oscillator and the elastic strip modes is given approximately by \cite{xu2021cavity}
\begin{align}
g_{1 (2), j} =& -\frac{\Omega_{b}}{2 d} \left(\frac{\hbar}{2m\omega_j}\right)^{1/2} \sin\left(\frac{\pi j }{ L}  \times \frac{L\mp D}{2} \right),
\label{couplingeq}
\end{align}
where we have taken the pointlike limit for the capacitors ($\Delta L \rightarrow 0)$ utilizing the fact that $\Delta L$ is assumed to be much smaller than the length $L$ of the strip; there is no ultraviolet (UV) divergence in such a limit in the determination of the quantum dynamics of the LC oscillator systems given below, which is a consequence of the effective one dimensional nature of the elastic strip \cite{xu2021cavity}. The coupling strength (\ref{couplingeq}) allows closed form analytical solutions for the quantum dynamics.

Supposing that the LC oscillators and the elastic strip  are prepared initially at $t=0$ in a product state with the latter in a thermal state, the time evolution of the reduced oscillators system density matrix expanded in the Fock state basis can be expressed as follows (for derivation details, see Supplementary Material \cite{SM}, which includes Refs. \cite{bose1999scheme, mancini1997ponderomotive}):
\begin{align}
&\rho_{n_1 n_2,n'_1 n'_2}(t) =  \exp \Big( -it \Omega_{b} (n_1 +n_2 - n_1' - n_2') \nn \\
&+i p_1( t) \big[(n_1 +n_1' +1)(n_1 - n_1') +  (n_2 +n_2' + 1)(n_2 - n_2') \big] \nn \\
&+ i p_2( t) \big[ 2 n_1 n_2 - 2 n_1'n_2' + n_1 + n_2 - n_1' - n_2'\big] \nn \\
&- d_1( t) \big[(n_1 - n_1')^2 +  (n_2 - n_2')^2\big] \nn \\
&-d_2(t) (n_1 - n_1')(n_2 - n_2')  \Big) \rho_{n_1 n_2,n'_1 n'_2}(0),
\label{LCdensity}
\end{align}
where the respective time-dependent terms are given by
\begin{subequations}
\begin{align}
p_1( t) =&\lambda \bigg( \frac{\pi^2 \tau}{6} -\tau \Re[\mathrm{Li}_2(-e^{i\sigma}) ] + \Im \Big[  \frac{1}{2} \mathrm{Li}_3 \left(-e^{i (\tau+\sigma)} \right)  \nn \\
&+\frac{1}{2} \mathrm{Li}_3 \left(-e^{i (\tau-\sigma)} \right)- \mathrm{Li}_3 (e^{i\tau}) \Big] \bigg),\label{p1exp}\\
p_2(t) =&\lambda \bigg( \frac{\pi^2 \tau}{12} +\tau \Re[\mathrm{Li}_2(e^{i\sigma}) ]  - \Im \Big[ \mathrm{Li}_3 (-e^{-i\tau}) \nn \\
&+ \frac{1}{2} \mathrm{Li}_3 \left(e^{i (\tau-\sigma)} \right) +\frac{1}{2} \mathrm{Li}_3 \left(e^{i (\tau+\sigma)} \right) \Big]  \bigg), 
\label{p2limit}\\
d_1(t) =& \sum_{j=1}^{\infty} \frac{1 - \cos (\omega_j t)}{\omega_j^2} g_{1,j }^2 \coth\left(\frac{\beta \hbar}{2} \omega_j \right),
\label{d1expresion}\\
d_2( t) =&2\sum_{j=1}^{\infty} \frac{1 - \cos( \omega_j t)}{\omega_j^2} g_{1,j }g_{2,j}\coth\left(\frac{\beta \hbar}{2} \omega_j \right),
\label{d2expression}
\end{align}
\end{subequations}
with $\beta^{-1}=k_BT$, where $k_B$ is Boltzmann's constant and $T$ is the elastic strip (environment) temperature. The dimensionless numerical constant $
\lambda=\frac{\Omega_{b}^2 \hbar}{16 d^2 m \omega_{1}^3}$, and $\mathrm{Li}_s(\cdot)$ is the polylogarithm function of order $s$. Note that  in the above expressions we have also introduced the following  notations for the dimensionless time: $\tau = \omega_1 t$, and for the scaled distance ratio: $\sigma = \pi D/L$. 

We now make several observations based on the form of Eq. (\ref{LCdensity}) about the LC oscillators' reduced system dynamics. Apart from the free evolution term, the $p_1(t)$ and $d_1(t)$ terms correspond to environment induced renormalization and dephasing respectively of the individual LC oscillators, while the $p_2(t)$ and $d_2(t)$ terms encode the effective environment induced mutual dynamics between the two LC oscillators. In particular, we have competing processes here where a non-zero mutual phase term $p_2(t)$ can render the LC oscillators' reduced density matrix to be entangled, while the real dephasing terms $d_1(t)$ and $d_2(t)$  serve to counteract  the entanglement generation. However, since both the $d_1(t)$ and $d_2(t)$ terms contain the oscillating factor $1 - \cos(\omega_j t)$, in which the harmonic mechanical mode frequencies are equally-spaced, these two terms completely vanish at times $t=2\pi j/\omega_1, j=0,1,2,\dots$. This periodic, full rephasing phenomenon is crucial for the formation of  entanglement as we will see below; in particular, it allows for periodic time windows in which to probe the generated entanglement. We note that this full rephasing phenomenon is a consequence of the one dimensional nature of the long elastic strip with uniformly spaced vibrational  modes; only partial rephasing will occur for two dimensional, elastic membranes that have non-uniformly spaced vibrational modes \cite{xu2021cavity}. 

A closer look at the $p_2(t)$ term also reveals that it enforces causality for the model; this can be seen by performing a partial trace over one of the LC oscillator subsystems and noticing that its influence on the other oscillator is only through the $p_2(t)$ term. Causality requires that the physical state of one LC circuit will not be changed by the presence of the other within the time that it takes for phonons to travel the separation distance between the two capacitors: $\Delta t = \frac{D}{v_{\mathrm{ph}}}= \sqrt{\frac{2m}{F L}}D$, where $v_{\mathrm{ph}} = \sqrt{\frac{FL}{2m}}$ is the phonon speed. Considering the following inequalities for $\tau$ and $\sigma$: $\tau< \sigma$ (corresponding to $t< \Delta t$) and $\sigma<\pi$ (corresponding to $D< L$), $p_2(t)$ in Eq.~(\ref{p2limit}) can be rewritten as a combination of  Bernoulli polynomials that are verified to vanish exactly, and become non-zero only when $t>\Delta t$.  We stress that such a causally consistent result can only be obtained by an exact, field theoretic treatment of the environment \cite{jonsson2014,benincasa2014quantum} (see Supplementary Material for further details \cite{SM}). 

{\it Zero temperature entanglement dynamics.---} We now discuss the entanglement dynamics of the model. Since the interaction Hamiltonian commutes with the system Hamiltonian, entanglement can be realized only if both LC circuits are initially in a Fock state superposition. For simplicity, we shall consider an initial ($t=0$) superposition of zero  and single photon states for each LC circuit: $|\psi(0) \rangle = \frac{1}{2}\left(|0\rangle_l +|1\rangle_l\right) \otimes \left(|0\rangle_r +|1\rangle_r\right)$. We furthermore assume as before for calculational convenience that the LC oscillators and strip are initially in a product state. The latter is equivalent to suddenly switching on the optomechanical interaction at $t=0$. While unphysical (the capacitive couplings are always `on'), such an assumption may be justified by supposing that the LC oscillator superposition states are prepared on a timescale that is much shorter than the phonon travel time between the two oscillators. 

We shall first focus on the zero temperature limit for the phonon field (corresponding to  the vacuum field state of the elastic strip). Despite the zero temperature limit being a challenge to realize given the presence of low frequency modes of the long strip, it allows analytical expressions for the dephasing terms (see Supplementary Material \cite{SM}), and yields important information about the competition between dephasing and entanglement generation.

\begin{figure}[h]
\begin{center}
\includegraphics[width=2.8in]{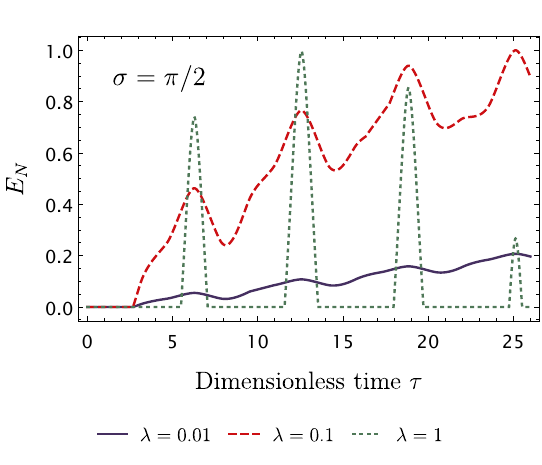} 
\caption{\label{figENzero} Logarithmic negativity plotted as a function of dimensionless time $\tau = \omega_1 t$ with different values of the numerical constant $\lambda$. The parameter $\sigma=\pi/2$ (corresponding to the LC circuits' separation $D=L/2$) }
\end{center}
\end{figure}
To determine whether the system is entangled, we utilize the logarithmic negativity $E_N(\rho)$ \cite{vidal2002computable} as our entanglement measure. 

With the full time evolution of the system density matrix given by Eq.~(\ref{LCdensity}) and the calculated time dependent terms, we obtain the logarithmic negativity $E_N$ as a function of the dimensionless time $\tau=\omega_1 t$ shown in Fig.~\ref{figENzero}; it can be seen  that the entanglement dynamics is sensitive to the value of the numerical constant $\lambda$, with several features in the time dependence noted as follows: (1) For the parameters considered here, the entanglement can only build up some time later than $t =\Delta t$ (corresponding to $\tau = \sigma$) in the `timelike' regime with respect to the phonon speed $v_{ph}$; this is a combined consequence of causality and the effect of zero temperature dephasing; although the environment induced phase term $p_2(t)$ starts to build up immediately after $t = \Delta t$, some additional time may be required in order to overcome the dephasing for entanglement to develop between the two subsystems. (2) $E_N$ is a local maximum at $\tau = 2 j \pi, ~j = 1,2,3,\dots$, corresponding to when both $d_1(t)$ and $d_2(t)$ vanish exactly, as noted previously.  Furthermore, depending on the value of the numerical constant $\lambda$, $E_N$ can get close to its upper bound value $=1$ for the two-level bipartite system, signaling a maximally entangled system state. (3) With the periodic vanishing of the dephasing terms, the maximally entangled state can always be generated regardless of the separation distance between the LC circuits; a larger separation distance only results in a longer time for the  entanglement to build up. 

{\it Entanglement dynamics with realistic conditions.---} We now turn our attention to more realistic scenarios where the strip has a finite temperature and both LC oscillators are subject to external dissipation, characterised by an assumed common decay rate $\kappa$. Considering the LC oscillators' frequency to be in the $\sim 10~{\mathrm{GHz}}$ regime with $\Omega_b\gg \kappa$, and the device temperature to be in the $\sim10~{\mathrm{mK}}$ regime, the dissipative quantum dynamics can be approximately described by the following zero temperature quantum master equation \cite{gardiner2004quantum}:
\begin{align}
\dot \rho_T =& -i[\mathcal{H}, \rho_T] + \sum_{k=1}^2 \kappa \mathcal{D}[a_k] \rho_T,
\label{mastereq}
\end{align}
where $\mathcal{H}$ is given by Eq. (\ref{Hamiltonian}), $\rho_T$ is the total density operator of LC oscillators and the phonon field, and the superoperator is given by $\mathcal{D}[O] \rho_T = O \rho_T O^\dag - \{O^\dag O, \rho_T\}/2 $. Assuming the same initial superposition state as previously for the LC oscillators and in the short time limit $\kappa t \ll 1$, the reduced density matrix of the two LC oscillators can be approximately obtained as the undamped solution Eq.~(\ref{LCdensity}) plus an external environment induced perturbation term $\rho_{\kappa}$, which is given by (for derivation details, see Supplementary Material \cite{SM})
\begin{align}
&\rho_{\kappa n_1 n_2, n_1' n_2'} = \kappa h(t) \Big [\sqrt{(n_1 +1)(n_1' +1)} \rho(t)_{n_1+1 n_2, n_1'+1 n_2'} \nn \\
&~~~+ \sqrt{(n_2 +1)(n_2' +1)} \rho(t)_{n_1 n_2+1, n_1' n_2'+1} \Big]- \frac{\kappa t}{2}(n_1 + n_1'  \nn \\
&~~~ + n_2 + n_2') \rho(t)_{n_1 n_2, n_1' n_2'}
\label{rhokappa}
\end{align}
where $h(t) =\int_0^t dt' e^{-2i(n_1 - n_1')p_1(t') - 2i (n_2 - n_2') p_2( t')} $, with $p_1(t)$ and $p_2(t)$ given by Eq.~(\ref{p1exp}) and Eq.~(\ref{p2limit}).

In the case of a finite temperature strip, the entanglement can be strongly suppressed due to the much more rapid thermal dephasing as compared with the zero temperature limit. However, the entanglement can nonetheless be present in the system around the times $\tau = 2 j \pi, j = 1,2,3.\dots$ when there is full rephasing. On the other hand, the dissipation of the LC oscillators due to their external environments will eventually destroy any possible entanglement in the system; we therefore shall only focus on the first peak of the entanglement when $\tau = 2 \pi$. In order to quantitatively investigate the entanglement dynamics, we assume some example parameters for the model that are related to actual experimental devices. In particular, for the elastic strip we adopt the silicon nitride vibrating string parameters from Ref. \cite{schilling2016near}: $\rho_m=10^3~{\mathrm{kg}}/{\mathrm{m}}^3$, $F=10^{-5}~{\mathrm{N}}$,  $W=1~\mu{\mathrm{m}}$, $T=0.1~\mu{\mathrm{m}}$; however, we assume a much longer length $L = 2~{\mathrm{cm}}$ than that considered in Ref. \cite{schilling2016near} ($\approx 60~\mu{\mathrm{m}}$) corresponding to the lowest mechanical mode frequency $\omega_1 \approx 50$ kHz. For the LC oscillators, we adopt typical superconducting microwave LC circuit parameters with $\Delta L=1~\mu{\mathrm{m}}$, $d=0.1~\mu{\mathrm{m}}$, and the circuit mode frequency of $\Omega/(2\pi)=15~{\mathrm{GHz}}$. The separation distance between the capacitors is taken to be $D = 1~\mathrm{cm}$.  We shall consider a decay rate constant $\kappa\approx 1~{\mathrm{kHz}}$; relaxation and dephasing times  ranging from a few hundred microseconds to over a millisecond have been reported for superconducting circuits \cite{reagor2016quantum,nersisyan_manufacturing_2019,wei2020verifying,zhang2021universal,somoroff2021millisecond}.  

\begin{figure}
\begin{center}
\includegraphics[width=3.3in]{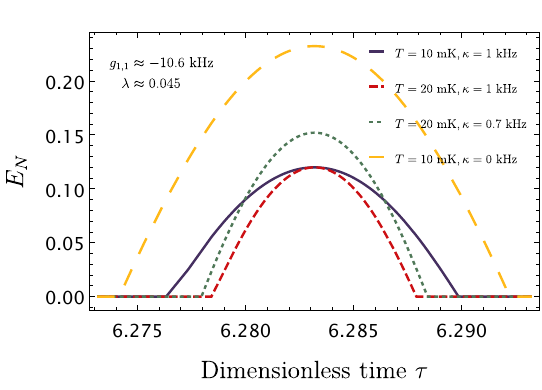}
\caption{\label{figENT} Logarithmic negativity plotted as a function of dimensionless time $\tau=\omega_1 t$ for a range of acoustic phonon field temperatures and LC oscillators' decay rates; the utilized parameters of the model discussed in the text correspond to $\lambda \approx 0.045$ and $g_{1,1} \approx -10.6$ kHz.}
\end{center}
\end{figure}

Using the above given parameters, we obtain the numerical results shown in Fig.~\ref{figENT} for the logarithmic negativity plotted around $\tau = 2\pi$, with a range of decay rates and  temperatures achievable in a dilution refrigerator. Note that the amount of entanglement at $\tau = 2 \pi$ when there is full rephasing [corresponding to $t\sim 126 ~\mu{\mathrm{s}}$, which is within the short time range to obtain the approximate solution Eq.~(\ref{rhokappa})] is not changed by the strip temperature (with the same decay constant). Instead, increasing the temperature narrows the time window (corresponding to a width around $250~\mathrm{ns}$ for $T=20~{\mathrm{mK}}$ in Fig. \ref{figENT}) 
during which the LC circuits system is entangled. On the other hand, increasing the decay rate causes both the entanglement maximum and  time window width to decrease. 

In order to experimentally probe the entanglement within the system, the initial and final LC systems' state may for example be prepared and measured by coupling the LC circuits to driven nonlinear Josephson phase qubits \cite{hofheinz2009synthesizing,blais2021circuit}. We also note that in the above analysis, we have ignored the loss channel due to the coupling of the acoustic strip modes to their external environments. This can be justified by noting that the quality ($Q$) factors of the low frequency mechanical modes may be engineered to have values as high as $Q=8 \times 10^8$ \cite{ghadimi2018elastic}, while the zero and single photon states of the LC oscillators induce strongly overlapping coherent states of the strip mechanical modes; such a decoherence channel can  be safely ignored on the relevant timescales for entanglement formation ($<1~{\mathrm{ms}}$).

{\it Conclusion.---}  We have investigated the entanglement dynamics of two LC oscillators coupled to a long elastic strip--a model system realization for two separated, localized UDW detectors interacting with a $1+1$ dimensional, massless scalar field. Exact solutions for the quantum time evolution of the oscillators were obtained, and the causality of the quantum dynamics analysed. 

With potential applications to quantum information processing in mind, it would be interesting to extend our model to multiple LC circuits and investigate possible multipartite entanglement generation via the optomechanical interaction \cite{bruschi2019} with a common, thermal acoustic environment, such as a long elastic strip or large surface area elastic membrane \cite{xu2021cavity}. It would also be interesting to come up with ways to increase the coupling between the strip and the capacitors, thereby leading to stronger signatures of entanglement; larger effective optomechanical couplings can be achieved for example through the placement of a Cooper-pair transistor between the LC oscillator and gated mechanical strip \cite{rimberg2014cavity}, or by engineering a strong LC oscillator-transmission line photon ``pressure" coupling for an all-microwave circuit realization \cite{johansson2014optomechanical,bothner2021photon}.

We thank Shadi Ali Ahmad, Sougato Bose, David Bruschi, Bei-Lok Hu, Eduardo Martin-Martinez, and Shih-Yuin Lin for very helpful discussions. This work is supported by a Dartmouth Teaching Fellowship and by the NSF under grant no. PHY-2011382.

\bibliography{main}
\end{document}


\title{Supplementary Material:  Optomechanical quantum entanglement mediated by  acoustic phonon fields}

\author{Qidong Xu}
\email[]{qidong.xu.gr@dartmouth.edu}

\author{M. P. Blencowe}
\email[]{miles.p.blencowe@dartmouth.edu}

\address{Department of Physics and Astronomy, Dartmouth College, Hanover, New Hampshire 03755, USA}

\date{\today}
\maketitle

\onecolumngrid

\section{Derivation for the reduced density matrix of two LC oscillators}
In this section we shall first provide a short derivation for the reduced density matrix of two LC oscillators without the environmental loss and then solve for the dynamics when the coupling between the LC oscillators and the external environment is taken into consideration.

Recall that the optomechanical Hamiltonian of the LC oscillators and the acoustic phonon environment is
\begin{align}
\mathcal{H}  = \sum_{k=1}^2 \Big[ \hbar \Omega_b \left(a_k^\dag a_k +\frac{1}{2}\right) +\sum^\infty_{j = 1} \hbar g_{k,j} \left(a_k^\dag a_k +\frac{1}{2}\right)   \left(b_j +b_j^\dag\right) \Big] + \sum^\infty_{j=1} \hbar \omega_j \left(b_j^\dag b_j +\frac{1}{2} \right).
\label{Hamiltonian}
\end{align}
A simpler Hamiltonian when there is only one LC oscillator has been discussed in \cite{xu2020toy}. Working in a basis of LC oscillators' number states and environment mode coherent states $|n_1,n_2,\{\alpha_i\}\rangle$, a straightforward generalization of the result in Ref.~(\cite{xu2020toy}) shows that
\begin{align}
e^{-\frac{i\mathcal{H} t}{\hbar}}|n_1,n_2, \{\alpha_i\} \rangle =& \exp\Bigg(-it\left[\Omega_b\left(n_1+ n_2 +1\right) + \sum_i \left(\frac{\omega_i}{2} - \frac{ \left[ g_{1,i}\left(n_1+\frac{1}{2}\right) + g_{2,i}\left(n_2+\frac{1}{2}\right) \right]^2}{\omega_i} \right) \right]\nn \\
&-\sum_i \frac{i\left[ g_{1,i}\left(n_1+\frac{1}{2}\right) + g_{2,i}\left(n_2+\frac{1}{2}\right) \right]^2}{\omega_i^2} \sin \omega_i t + \frac{1}{2}\sum_i \frac{ g_{1,i}\left(n_1+\frac{1}{2}\right) + g_{2,i}\left(n_2+\frac{1}{2}\right)}{\omega_i} \nn \\
&\times \left[\alpha_i^* \left(1-e^{i\omega_i t}\right) -\alpha_i \left(1-e^{-i \omega_i t}\right) \right]\Bigg) \left| n_1, n_2, \left\{\alpha_i e^{-i\omega_i t} - \frac{g_{1,i} \left(n_1+\frac{1}{2}\right) + g_{2,i}(n_2 + 
\frac{1}{2})}{\omega_i} \left(1-e^{-i\omega_i t}\right) \right\} \right\rangle.
\label{ncoheq}
\end{align}
Assuming the acoustic phonon environment is initially in a thermal state, its initial density matrix can be expanded in the coherent basis as \cite{bose1999scheme}:
\begin{align}
\rho_{\mathrm{bath}} =& \prod_ i \frac{1}{\pi\left(e^{\beta \hbar \omega_i} - 1\right)} \int d \alpha_i^2 \exp\Big(-|\alpha_i|^2 \left(e^{\beta \hbar \omega_i}-1\right) \Big)  |\alpha_i\rangle \langle \alpha_i|,
\label{rhothermeq}
\end{align}
where $\beta^{-1}=k_BT$, with $k_B$ Boltzmann's constant and $T$ the environment temperature.

Performing a partial trace over the acoustic environment degrees of freedom, we then obtain the reduced density matrix for the LC oscillators as
\begin{align}
\rho_{n_1 n_2,n'_1 n'_2}(t) =&  \exp \Bigg( -it \Omega_{b} (n_1 +n_2 - n_1' - n_2') 
+it \sum_i \frac{\left[g_{1,i}\left(n_1 + \frac{1}{2}\right) +g_{2,i}\left(n_2 + \frac{1}{2}\right) \right]^2 - \left[g_{1,i}\left(n_1' + \frac{1}{2}\right) +g_{2,i}\left(n_2' + \frac{1}{2}\right) \right]^2}{\omega_i} \nn \\
&\times\big[(n_1 +n_1' +1)(n_1 - n_1') +  (n_2 +n_2' + 1)(n_2 - n_2') \big] 
- i \sum_i \sin(\omega_i t) \big[ 2 n_1 n_2 - 2 n_1'n_2' + n_1 + n_2 - n_1' - n_2'\big]\nn \\
&\times \frac{\left[g_{1,i}\left(n_1 + \frac{1}{2}\right) +g_{2,i}\left(n_2 + \frac{1}{2}\right) \right]^2 - \left[g_{1,i}\left(n_1' + \frac{1}{2}\right) +g_{2,i}\left(n_2' + \frac{1}{2}\right) \right]^2}{\omega_i^2} - \sum_i (1 - \cos (\omega_i t)) \coth \left(\frac{\beta \hbar \omega_i}{2}\right)\nn \\
&\times \frac{ \left[ g_{1,n} (n_1 - n_1') + g_{2,n} (n_2 - n_2') \right]^2}{\omega_i^2}\Bigg) \rho_{n_1 n_2,n'_1 n'_2}(0).
\label{LCdensity}
\end{align}
Substituting in the expression of the coupling constant $g$, and summing over the induced phase terms, one can then obtain the expression of the reduced density matrix as in main text. 
In particular, the analytical expression for the decoherence terms as in Eq.~(6c) and Eq.~(6d) in the main text can be found in the zero temperature limit ($\beta \rightarrow +\infty$):
\begin{align}
\lim_{\beta \rightarrow +\infty}d_1(t) =&\lim_{\beta \rightarrow +\infty} \sum_{j=1}^{\infty} \frac{1 - \cos (\omega_j t)}{\omega_j^2} g_{1,j }^2 \coth\left(\frac{\beta \hbar}{2} \omega_j \right) \nn\\
=&\lambda \bigg( \Re \Big[ \frac{1}{2}\mathrm{Li}_3 \left(-e^{-i(\tau-\sigma)}\right) + \frac{1}{2}\mathrm{Li}_3 \left(-e^{i(\tau+\sigma)}\right) -\mathrm{Li}_3 \left(e^{i\tau}\right) -\mathrm{Li}_3 \left(-e^{i\sigma}\right)  +\zeta(3) \Big] \bigg),
\\
\lim_{\beta \rightarrow +\infty}d_2( t) =&2\lim_{\beta \rightarrow +\infty}\sum_{j=1}^{\infty} \frac{1 - \cos( \omega_j t)}{\omega_j^2} g_{1,j }g_{2,j}\coth\left(\frac{\beta \hbar}{2} \omega_j \right) \nn \\
=&2\lambda \bigg( \Re \Big[ \mathrm{Li}_3 \left(-e^{-i\tau}\right) +\mathrm{Li}_3 \left(e^{i\sigma}\right)  -\frac{1}{2}\mathrm{Li}_3 \left(e^{i(\tau-\sigma)}\right) - \frac{1}{2}\mathrm{Li}_3 \left(e^{i(\tau+\sigma)}\right) +\frac{3}{4}\zeta(3)   \Big] \bigg),
\end{align}
where $\zeta$ is the Euler–Riemann zeta function and the definitions for $\lambda$, $\tau$ and $\sigma$ are given in the main text. 

Next, we consider the quantum dynamics for the reduced density matrix when the couplings to external environments are considered; this then corresponds to the zero temperature quantum master equation:
\begin{align}
\dot \rho_T =& -i[H, \rho_T] + \chi(\rho(T)) = -i[H, \rho_T] +\frac{\kappa}{2}(2 a_1  \rho_T a_1^\dag  - a_1^\dag a_1 \rho_T - \rho_T a_1^\dag  a_1 ) 
+ \frac{\kappa}{2}(2 a_2 \rho_T  a_2^\dag -  a_2^\dag  a_2 \rho_T - \rho_T a_2^\dag a_2).
\label{mastereq}
\end{align}
Note that $\rho_T$ here represents for the density matrix of LC oscillators plus phonon field and we assume the same circuit mode decay rate $\kappa$. To solve for the above master equation, we shall follow the steps in Ref. \cite{mancini1997ponderomotive}. First we note that, in the absence of the LC circuit environment, the time evolution operator for the elastic strip and the circuits can be found as follows \cite{mancini1997ponderomotive}:
\begin{align}
U(t) =& e^{i \sum_j f(\omega_j t) \big[\frac{g_{1j}}{\omega_j}(n_1 + \frac{1}{2}) + \frac{g_{2j}}{\omega_j}(n_2 + \frac{1}{2})\big]^2} e^{-2i \sum_j \left(\frac{g_{1j}}{\omega_j}(n_1 + \frac{1}{2}) + \frac{g_{2j}}{\omega_j}(n_2 + \frac{1}{2})\right)x(t) \sin\omega_j t} e^{-iH_0 t},
\end{align}
where $f(x) = x - \sin x$, $n$ is the number operator, $x(t) = b_j^\dag e^{-i\frac{\omega_j t}{2}} +b_j e^{i\frac{\omega_j t}{2}}$ and $H_0$ is the free Hamiltonian of the strip and the LC  circuits.

Introducing a new density operator $R_T$ defined as $\rho_T = U R_T U^\dag$, the time evolution of $R_T$ is
\begin{align}
\dot R_T = U^\dag \chi(U R_T U^\dag ) U := \Tilde{\chi}(R_T),
\label{Req}
\end{align}
where $\chi$ is defined in Eq.~(\ref{mastereq}) and we also defined the superoperator $\Tilde{\chi}$ in the above equation. Writing the solution of Eq.~(\ref{Req}) in the form of $R_T = R_0 + Y$ with $\dot R_0 = 0$ (so that $R_0$ corresponds to the free evolution), we then have $\dot Y =\Tilde{\chi}(R_0 + Y) $. Assuming that $\kappa$ is small enough, we can then apply the first Born approximation $\Tilde{\chi}(R_0 + Y) \approx \Tilde{\chi}(R_0) $. This then leads to the solution: $\rho_T = \rho_{0T} + \rho_{\kappa T}$, where
\begin{align}
\rho_{0T}(t) = U(t) \rho_{T}(0) U(t)^\dag,
\end{align}
which corresponds to the time evolution in the absence of the external environment and 
\begin{align}
\rho_{\kappa T}(t) = U(t) \int_0^t \Tilde{\chi}(R_0, t') dt' U(t)^\dag.
\end{align}
After some algebra, the explicit expression for $\rho_{\kappa T}$ can be found as
\begin{align}
\rho_{\kappa T}(t) =& \kappa \int_0^t dt' e^{-2i \sum_j f(\omega_jt')\frac{g_{1j}}{\omega_j}N_j + 2i \sum_j  x(t')\frac{g_{1j}}{\omega_j}\sin\left[\frac{\omega_j}{2}t'\right]}a_1 \rho_{0T}(t) a_1^\dag e^{2i \sum_j f(\omega_jt')\frac{g_{1j}}{\omega_j}N_j - 2i \sum_j  x(t')\frac{g_{1j}}{\omega_j}\sin\left[\frac{\omega_j}{2}t'\right]} \nn \\
&- \frac{\kappa}{2}t a_1^\dag a_1 \rho_{0T}(t) - \frac{\kappa}{2}t \rho_{0T}(t) a_1^\dag a_1 +\{ 1 \rightarrow 2\},
\end{align}
where $N_j = \frac{g_{1j}}{\omega_j} n_1 + \frac{g_{2j}}{\omega_j} n_2$ and the notation $\{1\rightarrow2 \}$ means changing the subscription from 1 to 2 for the oscillator operators. With the initial superpotion of zero and single photon state for the LC oscillators, we note that the above expression is only valid when $\kappa t\ll 1$. To find the reduced density matrix for the LC circuits, we then perform a partial trace over the mechanical variables. The partial trace of $\rho_{0 T}(t)$ is just $\rho(t)$ as we have found in Eq.~(\ref{LCdensity}), and for the partial trace of $\rho_{\kappa T}$, we have
\begin{align}
\rho_{\kappa} :=\mathrm{Tr}_m \left[\rho_{\kappa T}\right] = \kappa \int_0^t dt' e^{-2i \sum_j f(\omega_jt')\frac{g_{1j}}{\omega_j}N_j}a_1 \rho(t) a_1^\dag e^{2i \sum_j f(\omega_jt')\frac{g_{1j}}{\omega_j}N_j} - \frac{\kappa}{2}t a_1^\dag a_1 \rho(t) - \frac{\kappa}{2}t \rho(t) a_1^\dag a_1 +\{ 1 \rightarrow 2\},
\end{align}
where we have used the cyclic property of the partial trace. Expanding in the Fock basis and summing over the mechanical modes, we obtain
\begin{align}
\rho_{\kappa n_1 n_2, n_1' n_2'} =& \kappa h(t) \Big [\sqrt{(n_1 +1)(n_1' +1)} \rho(t)_{n_1+1 n_2, n_1'+1 n_2'} + \sqrt{(n_2 +1)(n_2' +1)} \rho(t)_{n_1 n_2+1, n_1' n_2'+1} \Big]\nn \\
&- \frac{\kappa t}{2}(n_1 + n_1'  + n_2 + n_2') \rho(t)_{n_1 n_2, n_1' n_2'},
\end{align}
where $h(t) =\int_0^t dt' e^{-2i(n_1 - n_1')p_1(t') - 2i (n_2 - n_2') p_2( t')} $, with the expression of  $p_1(t)$ and $p_2(t)$ given in the main text.

\section{Causality analysis}
\begin{figure}[h]
\begin{center}
\includegraphics[width=4.0in]{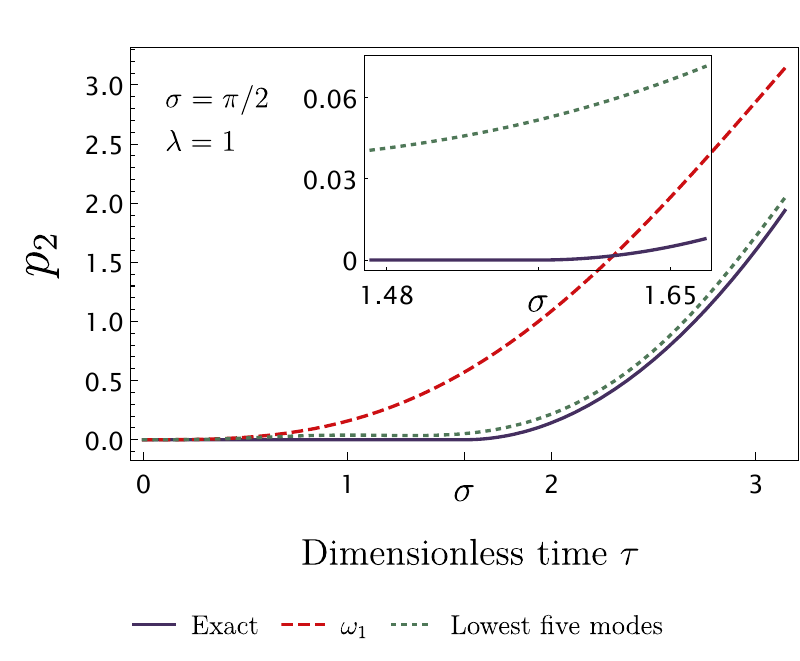} 
\caption{\label{figp2} The environment induced mutual phase term $p_2(t)$ plotted as a function of dimensionless time $\tau = \omega_1 t$. The constant $\lambda=1$ and $\sigma=\pi/2$ (corresponding to the LC circuits' separation $D=L/2$). Both the exact analytical expression (solid line) and finite mode sum approximations are shown for comparison: the contribution from the lowest, fundamental mode $\omega_1$ only (dashed line) and the contribution obtained by summing over the lowest five elastic frequency modes only (dotted line). The inset gives the zoomed in plot for $p_2$ close to $t= \sigma$.}
\end{center}
\end{figure}

The main purpose of this section is to show the importance of an exact field theoretic treatment to obtain a causal result  \cite{,jonsson2014,benincasa2014quantum}; by this we mean that one has to take account of the position-dependent coupling between system and the mechanical strip and sum over all mechanical degrees of freedom. If one approximately truncates to a finite number of field modes in the sum, or simply just considers the resonant mechanical mode as is common in optomechanical studies, causality is violated. For example, as we show in Fig.~\ref{figp2}, a strongly acausal result is obtained with only the contribution from the lowest, fundamental frequency mode of the elastic strip taken into account. By including more modes in the sum, the induced phase term $p_2(t)$  approaches its exact analytical expression, but nevertheless remains acausal.

\bibliography{smain}